\documentclass[11pt]{article}
\usepackage[tmargin=1in, lmargin=1.25in, rmargin=1.25in, bmargin=1.0in]{geometry}
\usepackage{setspace}
\usepackage[utf8]{inputenc}
\usepackage[backref,colorlinks,citecolor=blue,bookmarks=true]{hyperref}
\usepackage{mathtools, amssymb, amsthm}
\usepackage[capitalize]{cleveref}
\usepackage{enumerate}
\usepackage{todonotes}
\usepackage[affil-it]{authblk}

\theoremstyle{plain}
\newtheorem{theorem}{Theorem}[section]
\newtheorem{lemma}[theorem]{Lemma}
\newtheorem{corollary}[theorem]{Corollary}

\theoremstyle{definition}
\newtheorem{definition}[theorem]{Definition}

\theoremstyle{remark}

\newcommand*{\R}{{\mathbb{R}}}
\newcommand*{\cC}{{\mathcal{C}}}
\newcommand*{\cH}{{\mathcal{H}}}

\newcommand*{\cD}{{\mathcal{D}}}
\newcommand*{\cF}{{\mathcal{F}}}

\newcommand*{\cX}{{\mathcal{X}}}

\newcommand*{\cB}{{\mathcal{B}}}
\newcommand*{\cA}{{\mathcal{A}}}
\newcommand*{\cV}{{\mathcal{V}}}
\let\eps\epsilon
\DeclareMathOperator*{\pr}{\mathbb{P}}
\DeclareMathOperator*{\ex}{\mathbb{E}}
\DeclareMathOperator{\var}{Var}

\DeclareMathOperator{\tdeg}{deg_{\pm}}
\DeclareMathOperator{\adeg}{\widetilde{deg}}

\DeclarePairedDelimiter{\inn}{\langle}{\rangle}
\DeclarePairedDelimiter{\norm}{\|}{\|}

\newcommand{\dotp}[2]{#1 \boldsymbol{\cdot} #2}

\let\hat\widehat
\DeclareMathOperator{\poly}{poly}

\DeclareMathOperator{\sgn}{sign}
\newcommand{\opt}{\mathsf{opt}}
\newcommand{\etal}{\emph{et~al.}}
\newcommand{\acz}{\mathsf{AC}^0}

\title{The Polynomial Method is Universal for Distribution-Free  Correlational SQ Learning}
\author[1]{Aravind Gollakota}
\author[2]{Sushrut Karmalkar}
\author[1]{Adam Klivans}
\affil[1]{University of Texas at Austin}
\affil[2]{University of Wisconsin--Madison}
\date{October 22, 2020 \\ Revised August 24, 2023}

\begin{document}

\maketitle

\begin{abstract}

We consider the problem of distribution-free learning for Boolean function classes in the PAC and agnostic models. Generalizing a beautiful work of Malach and Shalev-Shwartz~\cite{malach2022hardness} that gave tight correlational SQ (CSQ) lower bounds for learning DNF formulas, we give new proofs that lower bounds on the threshold or approximate degree of any function class directly imply CSQ lower bounds for PAC or agnostic learning respectively. While such bounds implicitly follow by combining prior results by Feldman \cite{feldman2008evolvability,feldman2012complete} and Sherstov \cite{sherstov2008halfspace,sherstov2011pattern}, to our knowledge the precise statements we give had not appeared in this form before. Moreover, our proofs are simple and largely self-contained.

These lower bounds match corresponding positive results using upper bounds on the threshold or approximate degree in the SQ model for PAC or agnostic learning, and in this sense these results show that the polynomial method is a universal, best-possible approach for distribution-free CSQ learning.

\end{abstract}

\section{Introduction}

A successful and general approach to distribution-free learning in both Valiant's PAC model~\cite{valiant1984theory} and Kearns \etal's agnostic model~\cite{kearns1994toward} has been the so-called ``polynomial method.'' In this approach, a Boolean function is approximated by a low-degree polynomial with respect to various losses such as the $0/1$ loss of the sign of the polynomial or a uniform approximation by the polynomial itself.  This representation in turn leads to algorithms whose complexity is typically exponential in the degree of the approximating polynomial.  

Several researchers have pointed out that the polynomial method captures the best-known distribution-free learners for Boolean function classes  with the exception of classes that include parities \cite{hellerstein2007pac}. It is used for example in  the algorithm of Klivans and Servedio~\cite{klivans2004learning} for PAC-learning DNF formulas as well as the $L^1$ polynomial regression algorithm of Kalai \etal~\cite{kalai2008agnostically} for agnostically learning conjunctions. 

In this work we give a new proof characterizing (correlational) statistical query learning in the distribution-free PAC or agnostic setting in terms of the threshold degree or approximate degree of the unknown function class. To the best of our knowledge, such a characterization had previously only been implicit in work of \cite{feldman2008evolvability,feldman2012complete,sherstov2008halfspace,sherstov2011pattern}, as we discuss  shortly. This shows that unless we can find new algorithms that lie outside the SQ model, our only approach for distribution-free learning is to construct low-degree approximating polynomials. 

To be more concrete, let $f : \{-1, 1\}^n \to \{-1, 1\}$ be a Boolean function. The \emph{threshold degree} $\tdeg(f)$ of $f$ is the least degree of a polynomial $p$ such that $f(x) = \sgn(p(x))$ for all $x$. The pointwise \emph{$\epsilon$-approximate degree} $\adeg_\eps(f)$ of $f$ is the least degree of a polynomial $p$ such that $|f(x) - p(x)| \leq \epsilon$ for all $x$. It is well-known that for any class $\cC$ of functions with threshold degree bounded by $d$, $\cC$ can be viewed as a class of halfspaces over an $n^{O(d)}$-dimensional space (corresponding to the monomials up to degree $d$), and classic halfspace learners then yield distribution-free PAC learners for $\cC$ that run in $n^{O(d)}$ time. Similarly in the agnostic setting, the results of \cite{kalai2008agnostically} imply (as noted for example in \cite[Prop 2.1]{klivans2010lower}) that for a class $\cC$ with $\epsilon$-approximate degree bounded by $d$, degree-$d$ $L^1$ polynomial regression yields a distribution-free agnostic learner (with error $\opt + \epsilon$) running in time $n^{O(d)}$. This is since pointwise or $L^\infty$ approximation implies $L^2$ approximation (as required by \cite{kalai2008agnostically}) with respect to all distributions.

Our main results can now be stated as follows. Formal definitions may be found in the preliminaries.

\begin{theorem}\label{thm:tdeg-pac-intro}
Let $\cC$ be a Boolean function class satisfying some mild conditions (namely being closed under pattern restrictions, see \cref{def:pat-res}), with threshold degree $\Omega(d)$. Any distribution-free PAC learner for $\cC$ using only correlational statistical queries of tolerance $\tau < \frac{1}{10}$ requires at least $2^{\Omega(d)} \tau^2$ queries in order to learn $\cC$ up to error $\frac{1}{3}$. 
\end{theorem}

\begin{theorem}\label{thm:adeg-agnostic-intro}
Let $\cC$ be a Boolean function class as above (closed under pattern restrictions), with $\frac{1}{2}$-approximate degree $\Omega(d)$. Any distribution-free agnostic learner for $\cC$ using only correlational statistical queries of tolerance $\tau < \frac{1}{10}$ requires at least $2^{\Omega(d)} \tau^2$ queries in order to agnostically learn $\cC$ up to excess error $\frac{1}{100}$, i.e.\ true error $\opt + \frac{1}{100}$. 
\end{theorem}

Both these lower bounds match (up to logarithmic factors in the exponent) the upper bounds given by the polynomial method that were mentioned earlier.

As one example of a corollary, we obtain essentially tight lower bounds for the important problem of agnostically learning conjunctions (likewise disjunctions). Specifically, it is a classic result \cite{nisan1994degree, paturi1992degree} that the class of conjunctions on $\{-1, 1\}^n$ has $\epsilon$-approximate degree $\Theta(\sqrt{n})$ for all constant $\epsilon < 1$, and this yields a $2^{\Omega(\sqrt{n})}$ correlational SQ lower bound by the above theorem.

As indicated earlier, a slightly different characterization of CSQ PAC learnability in terms of the existence of low-weight threshold representations had been given in \cite{feldman2008evolvability} (see Theorem 5.4 therein). Combining this with work by Sherstov \cite{sherstov2008halfspace,sherstov2011pattern} on the relationships between various complexity measures of Boolean function classes as well as the links between learning and communication complexity, one could already obtain a qualitatively equivalent version of \cref{thm:tdeg-pac-intro}. Similarly but less directly, the ideas used to obtain \cref{thm:adeg-agnostic-intro} are arguably implicit in \cite{feldman2012complete} in combination with \cite{sherstov2008halfspace,sherstov2011pattern}. Thus the primary appeal of our results lies less in their novelty and more in the following aspects: (a) the statements are clean and explicitly reveal the dependence on both the degree $d$ and the tolerance $\tau$; (b) the proofs are direct, simple, and largely self-contained (without reference to communication complexity results); and (c) the proofs are more explicit about the hard instances for which the lower bound is obtained.

Both our theorems are based on a more general theorem that constructs a hard family of functions from a single function $f$ and a distribution $\mu$ which ``orthogonalizes'' it up to degree $d$, i.e.\ under which all degree-$d$ or lower moments of $f$ vanish. This condition has been used in prior works for proving lower bounds for learning neural networks with respect to Gaussian distributions (e.g.\ \cite{diakonikolas2020algorithms, diakonikolas2020near}).

Our main theorem crystallizes a basic principle of SQ lower bounds, saying that in the SQ setting, the hard distributions for a class of functions are the ones that orthogonalize them, i.e.\ zero out their low-degree moments:
\begin{theorem}
Let $f : \{-1, 1\}^{n/2} \to \R$ be a a Boolean function, and let $\mu$ be a distribution that orthogonalizes $f$ up to degree $d-1$. There exists a family $\cA$ of function-distribution pairs on $\{-1,1\}^n$ where each function in the family is of the form $x \mapsto f(x_V \oplus w)$ for some $V \subseteq [n]$ of size $n/2$ and $w \in \{-1, 1\}^{n/2}$, such that any distribution-free PAC learner for $\cA$ using only correlational statistical queries of tolerance $\tau < \frac{1}{10}$ requires at least $\Omega(2^{d} \tau^2)$ queries in order to learn $\cC$ up to error $\frac{1}{3}$. (Here $x_V$ denotes the vector of $x_i$ for $i \in V$, and $\oplus$ denotes the bitwise XOR.)
\end{theorem}
 
The proof is a simple application of the pattern matrix method due to Sherstov \cite{sherstov2011pattern}.   As a partial converse, it is not hard to see that distributions under which the target functions have significant low-degree moments allow Fourier-theoretic methods (e.g.\ the ``low degree algorithm'') to succeed. We view this result as providing an additional, more general sense in which the polynomial method is a ``universal'' approach to learning.

One unusual feature of our results is that rather than constructing a single hard distribution under which a large class of functions is hard to learn, our hard families involve a set of function-distribution pairs, where each function is effectively accompanied by its own hard distribution. A distribution-free learner should, of course, succeed equally well on such a family.

Our results also have consequences for the hardness of approximation by linear classes of classes with high threshold degree.
\begin{theorem}\label{thm:approx-hardness-intro}
Let $\cC$ be a Boolean function class closed under pattern restrictions, with threshold degree $\Omega(d)$. Consider the linear hypothesis class with an arbitrary $N$-dimensional embedding $\psi$ and a norm bound $B$: \[ \cH_{\psi, B} = \left\{ x \mapsto \dotp{\psi(x)}{w} \mid \norm{w}_2 \leq B \right\}. \] There exists a family $\cA$ of function-distribution pairs, with the functions lying in $\cC$, such that \[ \max_{(f, D) \in \cA} \ \min_{h \in \cH_{\psi, B}} \ \ex_{x \sim D}\left[\left(f(x) - h(x)\right)^2\right] > \frac{1}{2} - B \sqrt{N} 2^{-\Omega(d)}. \]
\end{theorem}
As one application, we see that for all classes with threshold degree $d = \omega(\log n)$ (which includes most classes, e.g.\ conjunctions, intersections of halfspaces, and $\acz$), so that $2^{d}$ is superpolynomial, even weak approximation is impossible unless $B\sqrt{N} = 2^{\Omega(d)}$. In particular, weak approximation is impossible with $B, N = \poly(n)$. This generalizes earlier results proved for classes such as conjunctions via the approximate rank \cite{klivans2010lower}.

Our approach builds closely on the recent work of \cite{malach2022hardness}, in which the authors introduce a new measure called the ``variance'' of a class of labeled distributions, and use it to show CSQ lower bounds as well as a host of hardness of approximation results for the class. In this paper we use the term ``correlational variance'' to refer to this quantity for clarity. They use this new framework to show an essentially-tight $2^{\Omega(n^{1/3})}$ CSQ lower bound for learning DNF formulas by applying technical results of Razborov and Sherstov \cite{razborov2010sign}. Our main contribution is to simplify their approach and extend this framework to the more general setting of orthogonalizing distributions.  We can then leverage known lower bounds on the threshold and approximate degree of function classes. For example, we can obtain the $2^{\Omega(n^{1/3})}$ lower bound of \cite{malach2022hardness} by using our main theorem with the classical $\Omega(n^{1/3})$ lower bound due to Minsky and Papert from the 60s \cite{minsky1969perceptrons}.  This deviates from established frameworks for proving CSQ lower bounds, which generally use variants of the SQ dimension (see \cite{feldman2016, reyzin2020statistical} for surveys).

\subsection{Related and prior work}
There has been a wealth of work on lower bounds for learning various Boolean function classes in various settings, and we can only hope to survey a small slice of it. All Boolean functions in what follow will be assumed to be on $n$ bits and $\poly(n)$-sized.

\paragraph{Cryptographic lower bounds} One of the original lines of work on the hardness of learning was based on cryptographic assumptions (see \cite[Chap.\ 6]{kearns1994introduction} for a textbook reference). Valiant in his original paper \cite{valiant1984theory} observed that pseudorandom families, and the function classes that can compute them, are inherently hard to learn in polynomial time. This was developed further by \cite{kearns1994cryptographic}, who proved hardness of polynomial-time learning for Boolean formulas and finite automata based on assumptions underlying public-key cryptography. Another important early work was \cite{kharitonov1993cryptographic}, which showed a quasipolynomial time lower bound for learning $\acz$ even under the uniform distribution (matching \cite{linial1993constant}) based on secure pseudorandom generators. For intersections of halfspaces, \cite{klivans2009cryptographic} showed the hardness of polynomial-time learning assuming the security of lattice-based cryptosystems.

\paragraph{Lower bounds on halfspace-based methods} Historically one of the chief approaches to learning various classes has been the use of halfspace-based learners such as Perceptron, SVM, and linear programming, including kernel methods and in particular the polynomial method. Measures such as approximate rank \cite{buhrman2001communication}, dimension and margin complexity \cite{vapnik2000nature,cristianini2000introduction, ben2002limitations,linial2007complexity} and their very recent probabilistic variants \cite{kamath2020approximate} are known to characterize inherent limitations of this approach. In \cite{klivans2010lower}, an approximate rank bound of $2^{\Omega(\sqrt{n})}$ was shown for disjunctions, and in \cite{razborov2010sign} a dimension complexity bound of $2^{\Omega(n^{1/3})}$ was shown for DNFs. Recent nearly-optimal lower bounds on threshold and approximate degree \cite{sherstov2013optimal,bun2019nearly,sherstov2019near} for classes such as intersections of halfspaces and $\acz$, as well as degree-weight tradeoffs \cite{servedio2012attribute}, may also be considered to fall into this line of work.

\paragraph{Complexity-theoretic lower bounds} In more recent times there has been a line of work \cite{daniely2014average, daniely2016complexity-hs, daniely2016complexity-dnf} establishing hardness of distribution-free learning using complexity theoretic assumptions such as the hardness of refuting random $k$-SAT. These works establish hardness of polynomial-time learning for DNF formulas and intersections of $\omega(\log n)$ halfspaces, and hardness of agnostic learning for conjunctions, halfspaces, and parities.

\paragraph{SQ lower bounds} The SQ model is the most general one in which unconditional lower bounds have been shown for many classes. The canonical result here is the SQ lower bound of $2^{\Omega(n)}$ for learning parities under the uniform distribution \cite{kearns1998efficient,blum1994weakly}, which introduced the SQ dimension. This measure has since been considerably generalized \cite{feldman2012complete, feldman2017statistical, feldman2017general}. As mentioned in the introduction, other important works in this overall line are those of \cite{feldman2008evolvability,feldman2012complete}, which provided characterizations and lower bounds for distribution-free CSQ learnability in both the PAC and agnostic settings. These can in turn be related to other commonly studied complexity measures using the far-reaching results of \cite{sherstov2008halfspace,sherstov2011pattern}. Our approach builds on the slightly different work of \cite{malach2022hardness}, which proved SQ lower bounds in terms of a measure called variance, variants of which had been considered before in \cite{shalev2017failures,shamir2018distribution}.

In the distribution-specific agnostic setting, the polynomial method takes the form of $L^1$ polynomial regression, introduced by \cite{kalai2008agnostically} as a universal approach to agnostic learning. The results of \cite{dachman2014approximate,diakonikolas2021optimality} later showed matching SQ lower bounds in terms of the $L^1$ approximate degree of a concept class, establishing this quantity as an appropriate measure of agnostic learnability (at least over the uniform distribution on the hypercube or the standard Gaussian). Our results may be seen as analogs of these results for distribution-free PAC and agnostic learnability.



\paragraph{SQ vs CSQ} 
The relationship between the general SQ model and the CSQ restriction has been a topic of some interest. In the distribution-specific setting, the two are unconditionally equivalent \cite{bshouty2002using}. More generally, the two are known to be equivalent when either the distribution is known, or sampling or nonuniformity is allowed \cite{feldman2008evolvability}. But somewhat surprisingly, the two are not in general equivalent in the distribution-free setting: \cite{feldman2008evolvability} showed that halfspaces are not efficiently CSQ-learnable even though they are always SQ-learnable. Thus our CSQ lower bounds do not readily extend to the general SQ model, and a very compelling direction for future work is to investigate whether such an extension is possible.

\paragraph{The polynomial method and known upper bounds} Essentially the only known distribution-free approach to learning most rich Boolean classes remains the polynomial method, i.e.\ representing Boolean functions as polynomial threshold functions (see \cite{hellerstein2007pac} for a survey) or, for agnostic learning, pointwise approximation (which in particular implies $L^1$ approximation \cite{kalai2008agnostically} with respect to any distribution). A notable exception in the perfectly realizable case is the use of (linear) algebraic methods for learning classes such as parities or more generally polynomials over finite fields \cite{hellerstein2007pac}. It is significant that the former category of methods may all be implemented using simple linear programming and made to work in the SQ setting, whereas the latter category is the prime (and essentially only) example of PAC algorithms that are not SQ. 

Improving upon the polynomial method has long been an elusive goal for many important classes. Indeed, in \cite{servedio2017circuit} the authors propose a change of perspective where rather than seek an efficient running time, the goal of a distribution-free learner is merely to run with nontrivial savings over exponential time, namely in time $2^{n - s(n)}$ where $s(n) = \omega(\log n)$. For $\acz$ circuits of size $M$ and depth $d$, they give such an algorithm with running time $2^{n - \Omega(n/(\log M)^{d-1})}$. 

\section{Preliminaries}

\paragraph{Notation} Let $f : \{-1, 1\}^n \to \{-1, 1\}$ be a Boolean function and $D$ be a distribution on $\{-1, 1\}^n$. We will sometimes view such functions as members of the $L^2$ space $L^2(\{-1, 1\}^n, D)$, with the inner product given by $\inn{f, g}_D = \ex_{D}[fg]$. When we use just $\inn{f, g}$ without any subscript, we will mean the inner product with respect to the uniform distribution on $\{-1, 1\}^n$. We denote by $f(D)$ the labeled distribution (on $\{-1, 1\}^n \times \{-1, 1\}$) of $(x, f(x))$ for $x \sim D$. We will generally denote unlabeled distributions on $\{-1, 1\}^n$ by non-calligraphic letters such as $D$ or the Greek $\mu$, and labeled distributions on $\{-1, 1\}^n \times \{-1, 1\}$ by calligraphic $\cD$ (unless it can be described as $f(D)$ for some $f, D$).

We use $[n]$ to denote $\{1, \dots, n\}$. Given $x \in \{-1, 1\}^n$ and $V \subseteq [n]$, we denote by $x_V$ the vector of $x_i$ for $i \in V$. Given $w \in \{-1, 1\}^n$, we denote by $x \oplus w$ the bitwise XOR (or in our case elementwise product, since we represent bits by $\{-1, 1\}$) of $x$ and $w$.

We also make use of basic notions from Boolean Fourier analysis. We use $\chi_S$ to denote the parity on $S \subseteq [n]$, $\chi_S(x) = \oplus_{i \in S} x_i = \prod_{i \in S} x_i$. The Fourier coefficients of a function $f : \{-1, 1\}^n \to \R$ are denoted $\hat{f}(S)$, with $\hat{f}(S) = \inn{f, \chi_S}$, and $f = \sum_{S \subseteq [n]} \hat{f}(S) \chi_S$.

\subsection{Learning in the statistical query model}
The statistical query (SQ) model was introduced by Kearns \cite{kearns1998efficient} and has proved highly influential as a realistic learning model that also allows strong lower bounds; see \cite{feldman2016, reyzin2020statistical} for surveys. Let $\cD$ be a labeled distribution on $\{-1, 1\}^n \times \{-1, 1\}$. An SQ oracle for $\cD$ is one that takes as input a query function $\phi : \{-1, 1\}^n \times \{-1, 1\} \to [-1, 1]$ and a tolerance $\tau \in (0, 1)$, and responds with a value in $[\ex_{\cD}[\phi] - \tau, \ex_{\cD}[\phi] + \tau]$. When a query is of the special form $\phi(x, y) = g(x) y$ for some $g : \{-1, 1\}^n \to [-1, 1]$, it is known as a correlational query, and is fully specified just by the function $g$. In the common case where $\cD$ is actually $f(D)$ for some function $f : \{-1, 1\}^n \to \{-1, 1\}$ and distribution $D$ on $\{-1, 1\}^n$, the expectation of a correlational query is $\ex_{(x, y) \sim f(D)}[g(x)y] = \ex_{x \sim D}[f(x) g(x)] = \inn{f, g}_D$, and for this reason such queries are also called inner product queries.

We recall the definitions of (realizable) PAC and agnostic learning as applicable in the SQ model. Let $\cC$ be a Boolean function class on $\{-1, 1\}^n$. In the traditional, realizable PAC setting, the learner is given SQ access to $c(D)$ for an unknown $c \in \cC$ and arbitrary distribution $D$ on $\{-1, 1\}^n$, and is said to learn $\cC$ up to error $\eps$ if it is able to output a function $h$ such that $\pr_{x \sim D}[h(x) \neq c(x)] \leq \eps$. The learner is called a distribution-free learner if it has this guarantee irrespective of what $D$ is. In the agnostic setting, the labels no longer arise from a function $c \in \cC$. Instead the learner is given SQ access to an arbitrary labeled distribution $\cD$ on $\{-1, 1\}^n \times \{-1, 1\}$, and the goal is to be competitive with the best-fitting classifier in $\cC$. Letting $\opt = \inf_{c \in \cC}\pr_{(x, y) \sim \cD}[c(x) \neq y]$, a learner is said to agnostically learn $\cC$ up to (excess) error $\eps$ if it is able to output a function $h$ such that $\pr_{(x, y) \sim \cD}[h(x) \neq y] \leq \opt + \eps$. Again, the learner is called distribution-free if this holds for all $\cD$. 

For distribution-specific learners of Boolean classes, it is a well-known observation \cite{bshouty2002using} that correlational SQs (CSQs) are equivalent to general SQs. This is not known to be the case in the distribution-free setting in which we operate, but it is known that a general SQ learner implies a CSQ learner when either sampling or nonuniformity is allowed \cite{feldman2008evolvability}.

\subsection{Correlational variance}
In \cite{malach2022hardness}, the authors introduce a measure called the ``variance'' of a function class, or more precisely a family of labeled distributions, and use it to show lower bounds both on CSQ learning as well as approximation of this class. We use the term ``correlational variance'' to refer to this quantity for clarity. The setting is as follows. Let $\cA$ be a family of function-distribution pairs $(f, D)$, where $f : \{-1,1\}^n \to \{-1, 1\}$ and $D$ is a distribution on $\{-1,1\}^n$. $\cA$ can also be seen as the family of the labeled distributions $f(D)$. Such a family will play exactly the same role as the notion of the ``assumption class'' from \cite{kearns1994toward}: namely, it is from an unknown member of this family that the learner receives labeled data (or SQ access to it).

\begin{definition}[Correlational Variance]
Let $\cA$ be a family of function-distribution pairs as above. For any $\phi : \cX \to [-1, 1]$, define \[ \var(\cA, \phi) = \ex_{(f, D) \sim \cA} \inn{f, \phi}_D^2 = \frac{1}{|\cA|} \sum_{(f, D) \in \cA} \inn{f, \phi}_D^2. \] Then the ``correlational variance'' of $\cA$ is defined as \[ \var(\cA) = \sup_{\|\phi\|_\infty \leq 1} \var(\cA, \phi). \]
\end{definition}

In words, the correlational variance of a family captures the maximum variance (really the second moment) of the correlation of a query function $\phi$ with a random member of the family.


This measure yields lower bounds on SQ learning using correlational queries. The core of the proof is the following appealingly simple Markov/Chebyshev-style lemma, which bounds the number of functions in a family that can be highly correlated with a query. We include it with a short proof so as to capture the essential function of the definition of correlational variance.

\begin{lemma}[\cite{malach2022hardness}, Lemma 6]\label{lem:var-sq-lemma}
Let $\cA$ be a family as above. Fix a query function $\phi : \{-1, 1\}^n \to \{-1, 1\}$ and a tolerance $\tau$. Let $\cA_\phi$ denote the function-distribution pairs with correlation at least $\tau$ with $\phi$, i.e.\ $\{ (f, D) \in \cA \mid |\inn{f, \phi}_D| \geq \tau \}$. Then $|\cA_\phi| \leq \frac{\var(\cA)}{\tau^2} |\cA|$.
\end{lemma}
\begin{proof}
What fraction of pairs $(f, D)$ in $\cA$ satisfy $\inn{f, \phi}_D^2 \geq \tau^2$? By a simple Markov/Chebyshev bound, this fraction is at most $\frac{\var(\cA, \phi)}{\tau^2} \leq \frac{\var(\cA)}{\tau^2}$.
\end{proof}

Roughly speaking, this means that an adversarial SQ oracle that responds with $0$ to every query only allows the learner to rule out at most $\frac{\var(\cA)}{\tau^2} |\cA|$ functions per query, essentially forcing even a weak learner to use $\Omega(\frac{\tau^2}{\var(\cA)})$ queries in total. Formally, \cite{malach2022hardness} show the following bound.

\begin{theorem}[\cite{malach2022hardness}, Theorem 7]\label{thm:var-sq-bound}
Let $\cA$ be a family of function-distribution pairs as above. Consider a learner given CSQ access to an unknown $(f, D) \in \cA$ with the goal of finding $h$ such that the 0-1 error $\pr_{x \sim D}[f(x) \neq h(x)]$ is small. Then any such learner making only correlational queries of tolerance $\tau$ must use $\Omega(\frac{\tau^2}{\var(\cA)})$ queries in order to output a function with 0-1 error better than $\frac{1}{2} - \frac{\tau}{2}$. In particular, assuming $\tau \leq \frac{1}{10}$ (say), $\Omega(\frac{\tau^2}{\var(\cA)})$ queries are required in order to obtain 0-1 error at most $\frac{1}{3}$.
\end{theorem}

The correlational variance of a family $\cA$ can be bounded in terms of the spectral norm of the linear operator $M(\cA)$ mapping $\phi : \cX \to [-1, 1]$ to the vector $(\inn{f, \phi}_D)_{(f, D) \in \cA}$. Here the domain $\cX$ will be $\{-1, 1\}^n$ for us. If we vectorize $\phi$ as $v(\phi) = (\phi(x))_{x \in \cX}$, then $M(\cA)$ is the matrix of size $|\cA| \times |\cX|$ whose rows are indexed by $(f, D) \in \cA$ and the columns by $x \in \cX$, and whose entries are given by $[f(x)D(x)]_{(f, D) \in \cA, x \in \cX}$. We then have the following bound.
\begin{lemma} \label{lem:var-spectral-norm}
\[ \var(\cA) \leq \frac{|\cX|}{|\cA|} \norm{M(\cA)}^2. \]
\end{lemma}
\begin{proof} This follows immediately from the observation that for any $\phi : \cX \to [-1, 1]$, \[ \var(\cA, \phi) = \frac{1}{|\cA|} \norm{M(\cA) v(\phi)}^2 \leq \frac{1}{|\cA|} \norm{M(\cA)}^2 \norm{v(\phi)}^2 \leq \frac{|\cX|}{|\cA|} \norm{M(\cA)}^2, \] where $\norm{v(\phi)}^2 \leq |\cX|$ since $\phi : \cX \to [-1, 1]$.
\end{proof}

Bounds on correlational variance also have implications for hardness of approximation, which we cover in \cref{sec:approx-hardness}.

\subsection{Orthogonalizing distributions}
The notion of an orthogonalizing distribution will be important to us, as it is the most general setting in which our results can be stated. It is a notion that has been used in many prior CSQ bounds and communication complexity results \cite{sherstov08communicationlower, sherstov2011pattern, razborov2010sign, diakonikolas2020algorithms, diakonikolas2020near}.

\begin{definition}[Orthogonalizing distribution]
Let $f : \{-1, 1\}^n \to \{-1, 1\}$ be a Boolean function. We say a distribution $\mu$ on $\{-1, 1\}^n$ orthogonalizes $f$ up to degree $d$ if for all polynomials of degree at most $d$, $\inn{f, p}_\mu = 0$. In particular, $\inn{f, \chi_S}_\mu = 0$ for all $|S| \leq d$.
\end{definition}

For us a convenient way to use this definition will be to define a function $g = f\mu$, and to observe that for all $|S| \leq d$, \[ \hat{g}(S) = \inn{g, \chi_S} = 2^{-n} \sum_{x \in  \{-1,1\}^n} f(x) \mu(x) \chi_S(x) = 2^{-n} \inn{f, \chi_S}_\mu = 0, \] as well as that for all $S \subseteq [n]$, since $|\inn{f, \chi_S}_\mu| \leq 1$, \[
    |\hat{g}(S)| = |\inn{g, \chi_S}| = 2^{-n} |\inn{f, \chi_S}_\mu| \leq 2^{-n}.
\]

\subsection{Threshold and approximate degree}

Two of the most basic ways of using real polynomials to represent a Boolean function are to represent it as a polynomial threshold function, and to approximate it pointwise. These are both classical notions that have a long history in approximation theory, and both notions are accompanied by a beautiful duality theory. In both cases the dual characterization can be viewed in terms of orthogonalizing distributions.

\begin{definition}[Threshold degree]
Let $f : \{-1,1\}^n \to \{-1,1\}$ be a function. The threshold degree of $f$, denoted $\tdeg(f)$, is the least degree of a real polynomial $p : \cX \to \R$ such that $f(x) = \sgn(p(x))$ for all $x$.
\end{definition}

It is a classical result that the threshold degree has a dual characterization in terms of an orthogonalizing distribution; see e.g.\ \cite[Theorem 3.3]{sherstov2011pattern}.
\begin{theorem} \label{thm:tdeg-duality}
Let $f : \{-1,1\}^n \to \{-1,1\}$ be a function with threshold degree $d$. Then there exists a distribution $\mu$ on $\cX$ that orthogonalizes $f$ up to degree $d-1$. 
\end{theorem}

\begin{definition}[Approximate degree]
Let $f : \{-1,1\}^n \to \{-1,1\}$ be a function. The pointwise $\epsilon$-approximate degree of $f$, denoted $\adeg_\epsilon(f)$, is the least degree of a real polynomial $p : \{-1,1\}^n \to \R$ such that $|f(x) - p(x)| \leq \epsilon$ for all $x$.
\end{definition}

Again, the following dual characterization is well-known; see e.g.\ \cite[Theorem 3.2]{sherstov2011pattern}.

\begin{theorem} \label{thm:adeg-duality}
Let $f : \{-1,1\}^n \to \{-1,1\}$ be a function with $\adeg_\eps(f) = d$. Then there exists a function $\psi : \{-1,1\}^n \to \R$ such that: \begin{enumerate}[(a)]
    \item $\sum_x \psi(x) f(x) > \epsilon$,
    \item $\sum_x |\psi(x)| = 1$,
    \item $\inn{\psi, p} = 0$ for all polynomials $p$ of degree less than $d$.
\end{enumerate} In particular, if we define a distribution $\mu(x) = |\psi(x)|$ and let $h(x) = \sgn(\psi(x))$, then we see that $\mu$ orthogonalizes $h$ up to degree $d-1$, and that $\inn{h, f}_\mu > \epsilon$.
\end{theorem}

\subsection{Pattern matrices}

Our constructions of hard families make use of Sherstov's pattern matrix method \cite{sherstov2011pattern}, along the lines of the construction of \cite{malach2022hardness} for DNFs.

\begin{definition}[Pattern matrix]
Let $f : \{-1, 1\}^k \to \R$ be a function, and let $n > k$ be a multiple of $k$. Let $\cV(n, k)$ be the family of all size-$k$ subsets $V \subset [n]$ of the following form: dividing $[n]$ into $k$ consecutive blocks of size $n/k$, $V$ consists of exactly one index from each block. (Notice that $|\cV(n, k)| = (n/k)^k$.) The $(n, k, f)$-pattern matrix is the matrix of size $2^n \times  (n/k)^k 2^k$ whose rows are indexed by $x \in \{-1,1\}^n$ and columns by pairs $(V, w) \in \cV(n, k) \times \{-1,1\}^k$, and whose elements are given by $f(x_V \oplus w)$.
\end{definition}

The following definitions are similar but will make many of our theorems easier to state.
\begin{definition}[Pattern matrix class]
Let $f : \{-1, 1\}^k \to \R$ be a function, and let $n > k$ be a multiple of $k$. The $(n, k, f)$-pattern matrix class is the class of all functions on $\{-1, 1\}^n$ of the form $x \mapsto f(x_V \oplus w)$ for some $(V, w) \in \cV(n, k) \times \{-1, 1\}^k$. We also sometimes use the term $(n, k, f)$-pattern matrix family for a family of function-distribution pairs where the functions involved are generated from $f$ in this way.
\end{definition}

\begin{definition}[Pattern restrictions]\label{def:pat-res}
Let $\cC = \bigcup_{n > 0} \cC_n$ be the union of some classes $\cC_n$ of Boolean functions on $\{-1, 1\}^n$. We say $\cC$ is closed under pattern restrictions if for any $k$, $n$ a multiple of $k$, and any $f \in \cC_k$, the function $x \mapsto f(x_V \oplus w)$ on $\{-1, 1\}^n$ lies in $\cC_n$ for any $V \subseteq [n]$ of size $k$ and $w \in \{-1, 1\}^k$. In the common case where $n$ is a small constant multiple of $k$, we will often be somewhat loose and not explicitly distinguish between $\cC_{k}$ and $\cC_n$ and just refer to $\cC$. Indeed, one can consider $\cC_{k}$ to effectively be a subset of $\cC_{n}$ using only some $k$ out of $n$ bits.
\end{definition}
Most common Boolean classes are closed under pattern restrictions. The main notable exceptions are monotone classes such as monotone conjunctions.

We will need the following bound on the spectral norm of a pattern matrix in terms of the Fourier coefficients of $f$.

\begin{theorem}[\cite{sherstov2011pattern}, Theorem 4.3]\label{thm:pattern-matrix-norm}
Let $f : \{-1, 1\}^k \to \R$, and let $A$ be its $(n, k, f)$-pattern matrix. Let $s = 2^n 2^k (\frac{n}{k})^k$ be the number of entries in $A$. Then \[ \norm{A} = \sqrt{s} \max_{S \subseteq [k]} \left\{ |\hat{f}(S)| \left(\frac{k}{n}\right)^{|S|/2} \right\}. \]
\end{theorem}

\section{Main theorem: correlational variance bounds via orthogonalizing distributions}
Here we state our main theorem in its most general setting, that of orthogonalizing distributions. The theorem is a strong generalization of \cite[Theorem 12]{malach2022hardness}, which proved such a result for the special case of DNF formulas, leveraging a powerful communication complexity result of Razborov and Sherstov \cite{razborov2010sign}. Our main contribution lies in noting that this proof does not require the full strength of \cite{razborov2010sign}, and actually holds in a considerably general setting.  Our lower bounds in terms of threshold and approximate degree follow as a result of this main theorem, since their dual characterizations furnish exactly the kinds of orthogonalizing distributions we need. The proof is a simple application of the pattern matrix method. It is worth noting that the functions in the hard family are all in fact juntas.

\begin{theorem}\label{thm:orth-var}
Let $n > k$ be a multiple of $k$. Let $f : \{-1, 1\}^k \to \R$ be a a Boolean function, and let $\mu$ be a distribution that orthogonalizes $f$ up to degree $d-1$, i.e.\ a distribution such that $\inn{f, p}_\mu = 0$ for all polynomials of degree less than $d$. There exists a $(n, k, f)$-pattern matrix family $\cA$ of function-distribution pairs on $\{-1, 1\}^n$ such that $\var(\cA) \leq (\frac{n}{k})^{-d}$.
\end{theorem}
\begin{proof}
For every $(V, w) \in \cV(n, k) \times \{-1, 1\}^k$, define $f_{V, w}(x) = f(x_V \oplus w)$ and $D_{V, w}(x) = 2^{k - n} \mu(x_V \oplus w)$. One can verify that $D_{V, w}$ is a valid distribution on $\{-1,1\}^n$: \begin{align*} \sum_{x \in \{-1,1\}^n} D_{V, w}(x) &= 2^{k - n} \sum_{x \in \{-1, 1\}^n} \mu(x_V \oplus w) \\
&= 2^{k - n} \sum_{z \in \{-1, 1\}^k} \sum_{\substack{x \in \{-1, 1\}^n : \\ x_V = z}} \mu(x_V \oplus w) \\
&= 2^{k - n} 2^{n - k} \sum_{z \in \{-1, 1\}^k} \mu(z \oplus w) \\
&= 1. \end{align*} Our family will be \[ \cA = \left\{ (f_{V, w}, D_{V, w}) \mid (V, w) \in \cV(n, k) \times \{-1, 1\}^k \right\}. \]

To analyze this, let $g = 2^{k - n} f \mu$, and observe that $M(\cA)$ is precisely the (transpose of the) $(n, k, g)$-pattern matrix. For all $|S| < d$, since $\inn{f, \chi_S}_\mu = 0$, i.e. $\sum_{x} f(x) \chi_S(x) \mu(x) = 0$, we have that $\hat{g}(S) = \inn{g, \chi_S} = 0$. Moreover, we know that for all $S$, \[ \hat{g}(S) = \inn{g, \chi_S} = 2^{-k} \sum_{x \in \{-1, 1\}^k} 2^{k - n} f(x) \mu(x) \chi_S(x) \leq  2^{-n} \sum_{x \in \{-1, 1\}^k} \mu(x) = 2^{-n}. \] Thus by \cref{thm:pattern-matrix-norm}, we have that \[ \norm{M(\cA)} \leq \sqrt{s} 2^{-n} \left(\frac{k}{n}\right)^{d/2}, \] where $s = 2^n 2^k (n/k)^k$. Note that $|\cA| = 2^k (n/k)^k = 2^{-n} s$. So finally by \cref{lem:var-spectral-norm}, we have \[ \var(\cA) \leq \frac{2^n}{|\cA|} \norm{M(\cA)}^2 \leq \frac{2^n}{2^{-n}s} s 2^{-2n} \left(\frac{k}{n}\right)^{d} = \left(\frac{k}{n}\right)^{d}. \]
\end{proof}

The following corollary is immediate from \cref{thm:var-sq-bound}.
\begin{corollary}\label{cor:orth-var-sq}
Let $\cA$ be the family defined above. Any SQ learner for $\cA$ making only correlational queries of tolerance $\tau < \frac{1}{10}$ must use $\Omega \left( (\frac{n}{k})^d \tau^2 \right)$ queries in order to output a function with 0-1 error at most $\frac{1}{3}$.
\end{corollary}

\section{CSQ lower bounds for PAC learning in terms of threshold degree}
Our main lower bound for PAC learning in terms of threshold degree is the following.
\begin{theorem}\label{thm:tdeg-sq}
Let $n > k$ be a multiple of $k$. Let $f : \{-1, 1\}^{k} \to \{-1,1\}$ be a function of threshold degree $d$. Let $\cF$ denote the $(n, k, f)$-pattern matrix class. Any distribution-free SQ learner for $\cF$ making only correlational queries of tolerance $\tau < \frac{1}{10}$ must use $\Omega \left( (\frac{n}{k})^d \tau^2 \right)$ queries in order to output a function with 0-1 error at most $\frac{1}{3}$.
\end{theorem}
\begin{proof}
Letting $\mu$ be the orthogonalizing distribution (on $\{-1,1\}^{k}$) guaranteed by the dual characterization (\cref{thm:tdeg-duality}), this theorem follows immediately by considering the family $\cA$ from \cref{thm:orth-var} and \cref{cor:orth-var-sq}, since $\var(\cA) \leq (\frac{n}{k})^{-d}$.
\end{proof}

We can now prove our original \cref{thm:tdeg-pac-intro} as stated.
\begin{proof}[Proof of \cref{thm:tdeg-pac-intro}]
Let $\cC$ be a class closed under pattern restrictions (\cref{def:pat-res}), with threshold degree $\Omega(d)$. Pick a function $f : \{-1, 1\}^{n/2} \to \{-1, 1\}$ in $\cC$ of threshold degree $\Omega(d)$. (Technically $f$ lies in the version of $\cC$, call it $\cC_{n/2}$, on $n/2$ bits. But as noted in \cref{def:pat-res}, one can effectively view $\cC_{n/2}$ as being part of $\cC$. The threshold degree bound remains $\Omega(d)$ asymptotically.) Now the class $\cF$ constructed above (\cref{thm:tdeg-sq}, with $k = n/2$) must satisfy $\cF \subseteq \cC$. The theorem follows.
\end{proof}

One application of this theorem, already proved as a special case in \cite{malach2022hardness}, is a $2^{\Omega(n^{1/3})}$ SQ lower bound for learning DNF formulas on $\{-1, 1\}^n$, proved by leveraging the well-known $\Omega(n^{1/3})$ Minksy--Papert threshold degree bound for DNFs \cite{minsky1969perceptrons}. This bound matches up to logarithmic factors in the exponent the algorithm of \cite{klivans2004learning} for this problem.

Another application is to the circuit class $\acz$, for which the following essentially-optimal threshold degree bound was recently shown.
\begin{theorem}[\cite{sherstov2019near}]
For any constant $\delta > 0$, there exists an $\acz$ circuit on $n$ bits with threshold degree $\Omega(n^{1 - \delta})$.
\end{theorem}
By \cref{thm:tdeg-sq}, we obtain as a consequence a CSQ lower bound of $2^{\Omega(n^{1 - \delta})}$ (for any constant $\delta > 0$) for learning $\acz$. This is essentially the strongest possible lower bound for distribution-free CSQ learning of $\acz$.

Another important application is to intersections of two halfspaces on $\{-1, 1\}^n$, for which Sherstov \cite{sherstov2013optimal} showed an optimal threshold degree bound of $\Omega(n)$. This yields a CSQ lower bound of $2^{\Omega(n)}$ for distribution-free learning of this class, which is again the strongest possible and improves considerably on the bound of \cite{klivans2007unconditional} of $2^{\Omega(\sqrt{n})}$ for intersections of $\sqrt{n}$ halfspaces.

Needless to say, further applications hold for all classes with known threshold degree bounds, including symmetric functions, decision trees, and functions with known lower bounds on sensitivity \cite{nisan1994degree, buhrman2002complexity, huang2019induced, karthikeyan2020resolution}.

It is also worth noting that if we treat $k$ as a free parameter, then \cref{thm:tdeg-sq} yields hard families of $k$-juntas. To our knowledge, the class of parities on $k$ bits, i.e.\ $\{\chi_S \mid S \subseteq [n], |S| \leq k\}$, was the only class for which an optimal, exponential SQ lower bound of $\sum_{i \leq k} \binom{n}{i} = n^{\Theta(k)}$ was known (under the uniform distribution), matching the algorithm of \cite{mossel2003learning} (itself only a polynomial improvement over brute force). We see that by picking $f : \{-1, 1\}^k \to \{-1, 1\}$ to be any function of threshold degree $\Omega(k)$, such as an intersection of two halfspaces \cite{sherstov2013optimal}, then we can construct a family $\cA$ of $k$-juntas with a nearly-optimal distribution-free CSQ lower bound of $(\frac{n}{k})^{\Omega(k)}$.


\section{CSQ lower bounds for agnostic learning in terms of approximate degree}
Our lower bound for agnostic learning is obtained by exploiting the dual characterization of approximate degree (\cref{thm:adeg-duality}), which turns out to be precisely suited to this task. This is done as follows. Let $f$ be a function with high $\frac{1}{2}$-approximate degree. First, since the dual function $h$ is accompanied by an orthogonalizing distribution $\mu$, we can use \cref{thm:orth-var} to generate a family from $h$ that is hard to learn. Next, since the dual function $h$ and the original $f$ are well-correlated, specifically $\inn{f, h}_\mu > \frac{1}{2}$, we can use an agnostic learner for a class generated from $f$ to yield a (weak) PAC learner for the family generated from $h$. Since the latter problem is hard, the original problem is as well.

\begin{lemma}\label{lem:adeg-pac-sq}
Let $n > k$ be a multiple of $k$. Let $f : \{-1, 1\}^{k} \to \{-1, 1\}$ be such that $\adeg_\eps(f) = d$, and let $h :  \{-1, 1\}^k \to \{-1, 1\}$ and the orthogonalizing distribution $\mu$ on $\{-1, 1\}^{k}$ be as given by its dual characterization (\cref{thm:adeg-duality}). There exists a $(n, k, h)$-pattern matrix family $\cA$ of function-distribution pairs on $\{-1, 1\}^n$ such that any SQ learner for $\cA$ making only correlational queries of tolerance $\tau < \frac{1}{10}$ must use $\Omega \left( (\frac{n}{k})^d \tau^2 \right)$ queries in order to output a function with 0-1 error at most $\frac{1}{3}$.
\end{lemma}
\begin{proof}
This is immediate from \cref{thm:orth-var} and \cref{cor:orth-var-sq}.
\end{proof}

This lets us prove an agnostic learning hardness result by reducing the problem of PAC learning $\cA$ to agnostically learning a similar pattern matrix family generated from $f$.

\begin{lemma}\label{lem:agnostic-pac-reduction}
Let $f : \{-1, 1\}^{k} \to \{-1, 1\}$ be such that $\adeg_{1 - \alpha}(f) = d$. Let $\cF$ denote the $(n, k, f)$-pattern matrix class. Then a distribution-free agnostic learner for $\cF$ capable of achieving 0-1 error $\opt + \epsilon$ yields a PAC learner capable of achieving 0-1 error $\alpha/2 + \epsilon$ for the family $\cA$ in \cref{lem:adeg-pac-sq}.
\end{lemma}
\begin{proof}
For brevity let $D = D_{V, w}$ and let $f', h' = f_{V, w}, h_{V, w}$. It is not hard to see that $\inn{f', h'}_D = \inn{f, h}_\mu$: \begin{align*}
    \inn{f', h'}_D &= \ex_{x \sim D}[f'(x) h'(x)] \\
    &= 2^{k - n} \sum_{x \in \{-1, 1\}^n} f(x_V \oplus w) h(x_V \oplus w) \mu(x_V \oplus w) \\
    &= 2^{k - n} \sum_{z \in \{-1, 1\}^k} \sum_{\substack{x \in \{-1, 1\}^n : \\ x_V = z}} f(x_V \oplus w) h(x_V \oplus w) \mu(x_V \oplus w) \\
    &= 2^{k - n} 2^{n - k} \sum_{z \in \{-1, 1\}^k} f(z \oplus w) h(z \oplus w) \mu(z \oplus w) \\
    &= \sum_{z \in \{-1, 1\}^k} f(z) h(z) \mu(z) \\
    &= \inn{f, h}_\mu.
\end{align*} So $\inn{f', h'}_D = \inn{f, h}_\mu > 1 - \alpha$, by \cref{thm:adeg-duality}. This means $\pr_{x \sim D}[f'(x) \neq h'(x)] = (1 - \inn{f', h'}_D)/2 < \alpha/2$. In other words, each function-distribution pair  $(h', D) \in \cA$ has $\opt = \inf_{f' \in \cF}\pr_{x \sim D}[f'(x) \neq h'(x)] < \alpha/2$ with respect to $\cF$. So if we could agnostically learn $\cF$ with error $\opt + \eps$ in a distribution-free way, then we could PAC-learn $\cA$ with error $\alpha/2 + \eps$. 
\end{proof}

\begin{theorem}\label{thm:adeg-agnostic-sq}
Let $n > k$ be a multiple of $k$. Let $f : \{-1, 1\}^{k} \to \{-1, 1\}$ be such that $\adeg_{1/2}(f) = d$. Let $\cF$ denote the $(n, k, f)$-pattern matrix class. Any distribution-free agnostic learner for $\cF$ using only correlational statistical queries of tolerance $\tau < \frac{1}{10}$ requires at least $\Omega \left( (\frac{n}{k})^d \tau^2 \right)$ queries in order to output a function with excess 0-1 error $\frac{1}{100}$, i.e.\ true 0-1 error $\opt + \frac{1}{100}$.
\end{theorem}
\begin{proof}
Suppose we had such a learner. By \cref{lem:agnostic-pac-reduction}, taking $\alpha = \frac{1}{2}$, this would be a PAC learner for $\cA$ capable of achieving 0-1 error at most $\frac{1}{4} + \frac{1}{100} < \frac{1}{3}$. Such a learner must obey the bound in \cref{lem:adeg-pac-sq}.
\end{proof}

\cref{thm:adeg-agnostic-intro} now follows in just the same way as \cref{thm:tdeg-pac-intro}.

An important application of this theorem is the problem of agnostically learning conjunctions (likewise disjunctions) in the distribution-free setting. This was one of the original problems considered in \cite{kearns1994toward}, and has seen hardness results in various restricted settings over the years, notable among which are a $2^{\Omega(\sqrt{n})}$ lower bound on Perceptron-based approaches (via the approximate rank) \cite{klivans2010lower} and a super-polynomial (but not exponential) CSQ lower bound for learning monotone conjunctions under the uniform distribution \cite{feldman2012complete}. Since the $\frac{1}{2}$-approximate degree of conjunctions on $n$ bits is $\Theta(\sqrt{n})$ \cite{nisan1994degree, paturi1992degree}, we obtain a CSQ lower bound of $2^{\Omega(\sqrt{n})}$ for this problem. This is essentially the best possible CSQ lower bound for this problem. 

Another important application is the problem of agnostically learning halfspaces \cite{kalai2008agnostically}. Halfspaces can compute Majority functions on $n$ bits, which have an approximate degree of $\Omega(n)$ \cite{paturi1992degree}. This yields a $2^{\Omega(n)}$ CSQ lower bound distribution-free agnostic learning of halfspaces, which is the strongest possible bound. Plugging in known approximate degree bounds for other classes such as symmetric Boolean functions \cite{paturi1992degree, buhrman2002complexity} yield further applications.

One can also apply this theorem to the class $\acz$, even though the resulting theorem is already implied by the PAC lower bound proved via threshold degree earlier. It is interesting that this bound can also be proved directly, via the following approximate degree bound.
\begin{theorem}[\cite{bun2019nearly}]
For any constant $\delta > 0$, there exists an $\acz$ circuit on $n$ bits with $\frac{1}{2}$-approximate degree $\Omega(n^{1 - \delta})$.
\end{theorem}
This yields a CSQ lower bound of $2^{\Omega(n^{1 - \delta})}$ (for any constant $\delta$) for distribution-free agnostic learning of $\acz$.

\subsection{CSQ lower bounds on attribute-efficient agnostic learning of sparse halfspaces}
The problem of attribute-efficient learning \cite{blum1990learning,blum1997selection} formalizes a notion of learning in the presence of irrelevant attributes, and is an important open problem in learning theory. Consider the class of $k$-sparse halfspaces on $\{-1, 1\}^n$. Since it has VC dimension $O(k \log n)$, from a statistical point of view a sample complexity of $\poly(k \log n)$ suffices to learn it (both in the PAC and agnostic settings). An efficient learner which achieves this sample complexity would be called an attribute-efficient learner. But despite years of research, no distribution-free attribute-efficient learners are known for this basic class.

In the SQ setting, the appropriate analog of sample complexity is the tolerance. Specifically, it takes $\Theta(1/\tau^2)$ samples to simulate a query of tolerance $\tau$, and this is sometimes known as the estimation complexity of an SQ algorithm using tolerance $\tau$. Accordingly, Feldman \cite{feldman2014open} posed the following question as the problem of attribute-efficient SQ learning of sparse halfspaces: does there exist an SQ algorithm capable of learning $k$-sparse halfspaces on $\{-1, 1\}^n$ only using $\poly(n)$ queries of tolerance $(k \log n)^{-O(1)}$? Though the question was originally asked in the PAC setting, here we answer it in the negative in  the distribution-free, agnostic, CSQ setting. The result follows readily from our agnostic learning lower bound in terms of approximate degree.

\begin{theorem}
There exists a family $\cF$ of $k$-sparse halfspaces on $\{-1, 1\}^n$ such that any distribution-free agnostic learner for $\cF$ only using correlational queries of tolerance $\tau < \frac{1}{10}$ requires at least $(\frac{n}{k})^{\Omega(k)}\tau^2$ queries in order to agnostically learn $\cF$ up to excess error $\frac{1}{10}$. In particular, for any $\tau = (k \log n)^{-O(1)}$, the lower bound remains $(\frac{n}{k})^{\Omega(k)}$ asymptotically, meaning no attribute-efficient CSQ learner exists.
\end{theorem}
\begin{proof}
Taking $f : \{-1, 1\}^k \to \{-1, 1\}$ to be a halfspace with $\frac{1}{2}$-approximate degree $\Omega(k)$, such as Majority, this is a direct application of \cref{thm:adeg-agnostic-sq}.
\end{proof}

\section{Hardness of approximation in terms of approximate degree}\label{sec:approx-hardness}
In addition to CSQ lower bounds, \cite{malach2022hardness} also prove hardness of approximation results for a family in terms of its correlational variance. Given a hypothesis class $\cH$ and a family $\cA$ of function-distribution pairs, a hardness of approximation result takes the form of a statement that in the worst case over a choice of $(f, D) \in \cA$, no hypothesis $h \in \cH$ can even achieve nontrivial loss with respect to $f(D)$, i.e.\ approximate $f$ wrt $D$. Naturally, this certainly implies that $\cH$ cannot be used to learn $\cA$. Our bounds on correlational variance in terms of approximate degree have consequences for hardness of approximation by kernelized linear functions, establishing inherent limitations of kernel methods for learning $\cA$. In particular, they show that the polynomial kernel is essentially an optimal kernel map in the distribution-free setting. This amounts to a variant of a result already observed in \cite{sherstov2011pattern}.

Let $f, h : \{-1, 1\}^n \to \R$ be functions, and let $D$ be a distribution on $\{-1, 1\}^n$. Let $L_{f(D)}(h)$ denote the squared loss $\ex_{x \sim D}[(h(x) - f(x))^2] = \|h - f\|_D^2$. Fix any embedding (or kernel feature map) $\psi : \{-1, 1\}^n \to [-1, 1]^N$ for any $N$, $B > 0$, and define the kernelized linear class \[ \cH_{\psi, B} = \left\{ x \mapsto \dotp{\psi(x)}{w} \mid \norm{w}_2 \leq B \right\}. \] Then the following bound on approximation by $\cH_{\psi, B}$ holds.

\begin{theorem}[\cite{malach2022hardness}]
Let $\cA$ be a family of function-distribution pairs on $\{-1, 1\}^n$. Then \[ \max_{(f, D) \in \cA} \ \min_{h \in \cH_{\psi, B}} \ L_{f(D)}(h) \ \geq \ex_{(f, D) \sim \cA} \ \min_{h \in \cH_{\psi, B}} \ L_{f(D)}(h) \ > \ \frac{1}{2} - B \sqrt{N} \sqrt{\var(\cA)}. \]
\end{theorem}

A similar but more involved result can also be proved for approximation by depth-two neural networks; see \cite[Theorem 3]{malach2022hardness}.

We obtain the following consequence.
\begin{theorem}
Let $\cC$ be a Boolean function class closed under pattern restrictions, with $(1 - \alpha)$-approximate degree $d$, where $\alpha$ is a sufficiently small constant ($\alpha = 1/16$ suffices). Let $\cH_{\psi,B}$ be the linear hypothesis class defined above. Then there exists a family $\cA$ of function-distribution pairs, with the functions lying in $\cC$, such that \[ \max_{(f, D) \in \cA} \ \min_{h \in \cH_{\psi, B}} \ L_{f(D)}(h) \ \geq \ex_{(f, D) \sim \cA} \ \min_{h \in \cH_{\psi, B}} \ L_{f(D)}(h) \ > \ \frac{1}{8} - B \sqrt{N} 2^{-\Omega(d)}. \]
\end{theorem}
\begin{proof}
Let $f : \{-1, 1\}^{n/2} \to \{-1, 1\}$ be a function in $\cC$ with $(1 - \alpha)$-approximate degree $d$. Let $g : \{-1, 1\}^{n/2} \to \{-1, 1\}$ and $\mu$ on $\{-1, 1\}^{n/2}$ be the accompanying dual function and distribution respectively, so that $\inn{f, g}_\mu \geq 1 - \alpha$.

Let $\cA$ denote the $(n, n/2, f)$-pattern matrix family of function-distribution pairs arising from $f$ and $\mu$, i.e.\ pairs of the form $(f_{V, w}, D_{V, w})$, where $f_{V, w}(x) = f(x_V \oplus w)$ and $D_{V, w}(x) = 2^{-n/2}\mu(x_V \oplus w)$. Similarly let $\cB$ denote the $(n, n/2, g)$-pattern matrix arising from $g$ and $\mu$. Recall that $\var(\cB) \leq 2^{-d}$, since $\mu$ orthogonalizes $g$ up to degree $d$.

Fix any $V, w$, and let $f' = f_{V, w}, g' = g_{V, w}, D' = D_{V, w}$ for brevity. It is easily calculated that $\inn{f', g'}_{D'} = \inn{f, g}_\mu \geq 1 - \alpha$, so that $\|f' - g'\|_{D'}^2 = 2 - 2\inn{f', g'}_{D'} \leq 2\alpha$. Now fix any $h \in \cH_{\psi, B}$. We will argue that if $h$ can approximate $f'$ under $D'$, then it can also approximate $g'$, simply because $f'$ and $g'$ are close under $D'$. Indeed, \begin{align*} L_{g'(D')}(h) &= \norm{g' - h}_{D'}^2 \\
&\leq 2\norm{f' - h}_{D'}^2 + 2\norm{f' - g'}_{D'}^2 \\
&\leq 2L_{f'(D')}(h) + 4\alpha,
\end{align*} i.e., \[ L_{f'(D')}(h) \geq \frac{1}{2} L_{g'(D')}(h) - 2\alpha. \] Thus for every $(f', D') \in \cA$, there is a corresponding $(g', D') \in \cB$ such that \[ \min_{h \in \cH_{\psi, B}} L_{f'(D')}(h) \geq \frac{1}{2} \min_{h \in \cH_{\psi, B}} L_{g'(D')}(h) - 2\alpha, \] and in fact this correspondence is one-to-one. So, taking the average, \begin{align*}
    \ex_{(f', D') \sim \cA} \ \min_{h \in \cH_{\psi, B}} \ L_{f'(D')}(h) &\geq \frac{1}{2} \ex_{(g', D') \sim \cB} \ \min_{h \in \cH_{\psi, B}} \ L_{g'(D')}(h) - 2\alpha \\
    &> \frac{1}{4} - \frac{B}{2} \sqrt{N} \sqrt{\var(\cB)} - 2\alpha \\
    &\geq \frac{1}{8} - \frac{B}{2} \sqrt{N} 2^{-d}
\end{align*} for $\alpha \leq 1/16$.
\end{proof}

Thus we see that for any class $\cC$ with approximate degree that is even super-logarithimic, i.e.\ $\omega(\log n)$, so that $2^{-\Omega(d)}$ is negligible, i.e.\ $n^{-\omega(1)}$, then no kernelized linear function (with $B$, $N$ being $\poly(n)$) can even weakly approximate $\cC$. If $B$ is fixed, then the dimension $N$ of the embedding must be taken to be $2^{\Omega(d)}$ for approximation to even be possible. This is essentially tight, since $d$ is of course the approximate degree, and so approximation is certainly possible using the polynomial kernel up to degree $d$ (so that $N = n^{O(d)}$). Thus in this sense the polynomial kernel is nearly optimal for distribution-free approximation.

\section*{Acknowledgements}
We thank Justin Thaler for generously sharing and discussing relevant references, as well as Eran Malach for helpful conversations. This work was done at UT Austin and partially supported by the NSF AI Institute for Foundations of Machine Learning (IFML).

\bibliography{refs}

\newcommand{\etalchar}[1]{$^{#1}$}
\begin{thebibliography}{DSFT{\etalchar{+}}14}

\bibitem[BDES02]{ben2002limitations}
Shai Ben-David, Nadav Eiron, and Hans~Ulrich Simon.
\newblock Limitations of learning via embeddings in euclidean half spaces.
\newblock {\em Journal of Machine Learning Research}, 3(Nov):441--461, 2002.

\bibitem[BdW01]{buhrman2001communication}
Harry Buhrman and Ronald de~Wolf.
\newblock Communication complexity lower bounds by polynomials.
\newblock In {\em Proceedings 16th Annual IEEE Conference on Computational
  Complexity}, pages 120--130. IEEE, 2001.

\bibitem[BDW02]{buhrman2002complexity}
Harry Buhrman and Ronald De~Wolf.
\newblock Complexity measures and decision tree complexity: a survey.
\newblock {\em Theoretical Computer Science}, 288(1):21--43, 2002.

\bibitem[BF02]{bshouty2002using}
Nader~H Bshouty and Vitaly Feldman.
\newblock On using extended statistical queries to avoid membership queries.
\newblock {\em Journal of Machine Learning Research}, 2(Feb):359--395, 2002.

\bibitem[BFJ{\etalchar{+}}94]{blum1994weakly}
Avrim Blum, Merrick Furst, Jeffrey Jackson, Michael Kearns, Yishay Mansour, and
  Steven Rudich.
\newblock Weakly learning dnf and characterizing statistical query learning
  using fourier analysis.
\newblock In {\em Proceedings of the twenty-sixth annual ACM symposium on
  Theory of computing}, pages 253--262, 1994.

\bibitem[BL97]{blum1997selection}
Avrim~L Blum and Pat Langley.
\newblock Selection of relevant features and examples in machine learning.
\newblock {\em Artificial intelligence}, 97(1-2):245--271, 1997.

\bibitem[Blu90]{blum1990learning}
Avrim Blum.
\newblock Learning boolean functions in an infinite attribute space.
\newblock In {\em Proceedings of the twenty-second annual ACM symposium on
  Theory of computing}, pages 64--72, 1990.

\bibitem[BT19]{bun2019nearly}
Mark Bun and Justin Thaler.
\newblock {A Nearly Optimal Lower Bound on the Approximate Degree of AC$^0$}.
\newblock {\em SIAM Journal on Computing}, 49(4):FOCS17--59--FOCS17--96, 2019.

\bibitem[CST{\etalchar{+}}00]{cristianini2000introduction}
Nello Cristianini, John Shawe-Taylor, et~al.
\newblock {\em An introduction to support vector machines and other
  kernel-based learning methods}.
\newblock Cambridge university press, 2000.

\bibitem[Dan16]{daniely2016complexity-hs}
Amit Daniely.
\newblock Complexity theoretic limitations on learning halfspaces.
\newblock In {\em Proceedings of the forty-eighth annual ACM symposium on
  Theory of Computing}, pages 105--117, 2016.

\bibitem[DKKZ20]{diakonikolas2020algorithms}
Ilias Diakonikolas, Daniel~M Kane, Vasilis Kontonis, and Nikos Zarifis.
\newblock Algorithms and sq lower bounds for pac learning one-hidden-layer relu
  networks.
\newblock In {\em Conference on Learning Theory}, pages 1514--1539, 2020.

\bibitem[DKPZ21]{diakonikolas2021optimality}
Ilias Diakonikolas, Daniel~M Kane, Thanasis Pittas, and Nikos Zarifis.
\newblock The optimality of polynomial regression for agnostic learning under
  gaussian marginals in the sq model.
\newblock In {\em Conference on Learning Theory}, pages 1552--1584. PMLR, 2021.

\bibitem[DKZ20]{diakonikolas2020near}
Ilias Diakonikolas, Daniel~M Kane, and Nikos Zarifis.
\newblock Near-optimal sq lower bounds for agnostically learning halfspaces and
  relus under gaussian marginals.
\newblock {\em arXiv preprint arXiv:2006.16200}, 2020.

\bibitem[DLS14]{daniely2014average}
Amit Daniely, Nati Linial, and Shai {Shalev-Shwartz}.
\newblock From average case complexity to improper learning complexity.
\newblock In {\em Proceedings of the forty-sixth annual ACM symposium on Theory
  of computing}, pages 441--448, 2014.

\bibitem[DS16]{daniely2016complexity-dnf}
Amit Daniely and Shai {Shalev-Shwartz}.
\newblock Complexity theoretic limitations on learning dnf’s.
\newblock In {\em Conference on Learning Theory}, pages 815--830, 2016.

\bibitem[DSFT{\etalchar{+}}14]{dachman2014approximate}
Dana Dachman-Soled, Vitaly Feldman, Li-Yang Tan, Andrew Wan, and Karl Wimmer.
\newblock Approximate resilience, monotonicity, and the complexity of agnostic
  learning.
\newblock In {\em Proceedings of the twenty-sixth annual ACM-SIAM symposium on
  Discrete algorithms}, pages 498--511. SIAM, 2014.

\bibitem[Fel08]{feldman2008evolvability}
Vitaly Feldman.
\newblock Evolvability from learning algorithms.
\newblock In {\em Proceedings of the fortieth annual ACM symposium on Theory of
  computing}, pages 619--628, 2008.

\bibitem[Fel12]{feldman2012complete}
Vitaly Feldman.
\newblock A complete characterization of statistical query learning with
  applications to evolvability.
\newblock {\em Journal of Computer and System Sciences}, 78(5):1444--1459,
  2012.

\bibitem[Fel14]{feldman2014open}
Vitaly Feldman.
\newblock Open problem: The statistical query complexity of learning sparse
  halfspaces.
\newblock In {\em Conference on Learning Theory}, pages 1283--1289, 2014.

\bibitem[Fel16]{feldman2016}
Vitaly Feldman.
\newblock Statistical query learning.
\newblock In Ming-Yang Kao, editor, {\em Encyclopedia of Algorithms}, pages
  2090--2095. Springer New York, New York, NY, 2016.

\bibitem[Fel17]{feldman2017general}
Vitaly Feldman.
\newblock A general characterization of the statistical query complexity.
\newblock In {\em Conference on Learning Theory}, pages 785--830. PMLR, 2017.

\bibitem[FGR{\etalchar{+}}17]{feldman2017statistical}
Vitaly Feldman, Elena Grigorescu, Lev Reyzin, Santosh~S Vempala, and Ying Xiao.
\newblock Statistical algorithms and a lower bound for detecting planted
  cliques.
\newblock {\em Journal of the ACM (JACM)}, 64(2):1--37, 2017.

\bibitem[HS07]{hellerstein2007pac}
Lisa Hellerstein and Rocco~A Servedio.
\newblock On pac learning algorithms for rich boolean function classes.
\newblock {\em Theoretical Computer Science}, 384(1):66--76, 2007.

\bibitem[Hua19]{huang2019induced}
Hao Huang.
\newblock Induced subgraphs of hypercubes and a proof of the sensitivity
  conjecture.
\newblock {\em Annals of Mathematics}, 190(3):949--955, 2019.

\bibitem[Kea98]{kearns1998efficient}
Michael Kearns.
\newblock Efficient noise-tolerant learning from statistical queries.
\newblock {\em Journal of the ACM (JACM)}, 45(6):983--1006, 1998.

\bibitem[Kha93]{kharitonov1993cryptographic}
Michael Kharitonov.
\newblock Cryptographic hardness of distribution-specific learning.
\newblock In {\em Proceedings of the twenty-fifth annual ACM symposium on
  Theory of computing}, pages 372--381, 1993.

\bibitem[KKMS08]{kalai2008agnostically}
Adam~Tauman Kalai, Adam~R Klivans, Yishay Mansour, and Rocco~A Servedio.
\newblock Agnostically learning halfspaces.
\newblock {\em SIAM Journal on Computing}, 37(6):1777--1805, 2008.

\bibitem[KMS20]{kamath2020approximate}
Pritish Kamath, Omar Montasser, and Nathan Srebro.
\newblock Approximate is good enough: Probabilistic variants of dimensional and
  margin complexity.
\newblock In {\em Conference on Learning Theory}, 2020.

\bibitem[KS04]{klivans2004learning}
Adam~R Klivans and Rocco~A Servedio.
\newblock Learning {DNF} in time $2^{O(n^{1/3})}$.
\newblock {\em Journal of Computer and System Sciences}, 68(2):303--318, 2004.

\bibitem[KS07]{klivans2007unconditional}
Adam~R Klivans and Alexander~A Sherstov.
\newblock Unconditional lower bounds for learning intersections of halfspaces.
\newblock {\em Machine Learning}, 69(2-3):97--114, 2007.

\bibitem[KS09]{klivans2009cryptographic}
Adam~R Klivans and Alexander~A Sherstov.
\newblock Cryptographic hardness for learning intersections of halfspaces.
\newblock {\em Journal of Computer and System Sciences}, 75(1):2--12, 2009.

\bibitem[KS10]{klivans2010lower}
Adam~R Klivans and Alexander~A Sherstov.
\newblock Lower bounds for agnostic learning via approximate rank.
\newblock {\em Computational Complexity}, 19(4):581--604, 2010.

\bibitem[KSP20]{karthikeyan2020resolution}
Rohan Karthikeyan, Siddharth Sinha, and Vallabh Patil.
\newblock On the resolution of the sensitivity conjecture.
\newblock {\em Bulletin of the American Mathematical Society}, 2020.

\bibitem[KSS94]{kearns1994toward}
Michael~J Kearns, Robert~E Schapire, and Linda~M Sellie.
\newblock Toward efficient agnostic learning.
\newblock {\em Machine Learning}, 17(2-3):115--141, 1994.

\bibitem[KV94a]{kearns1994cryptographic}
Michael Kearns and Leslie Valiant.
\newblock Cryptographic limitations on learning boolean formulae and finite
  automata.
\newblock {\em Journal of the ACM (JACM)}, 41(1):67--95, 1994.

\bibitem[KV94b]{kearns1994introduction}
Michael~J Kearns and Umesh Vazirani.
\newblock {\em An introduction to computational learning theory}.
\newblock MIT press, 1994.

\bibitem[LMN93]{linial1993constant}
Nathan Linial, Yishay Mansour, and Noam Nisan.
\newblock {Constant depth circuits, Fourier transform, and learnability}.
\newblock {\em Journal of the ACM (JACM)}, 40(3):607--620, 1993.

\bibitem[LMSS07]{linial2007complexity}
Nati Linial, Shahar Mendelson, Gideon Schechtman, and Adi Shraibman.
\newblock Complexity measures of sign matrices.
\newblock {\em Combinatorica}, 27(4):439--463, 2007.

\bibitem[MOS03]{mossel2003learning}
Elchanan Mossel, Ryan O'Donnell, and Rocco~P Servedio.
\newblock Learning juntas.
\newblock In {\em Proceedings of the thirty-fifth annual ACM symposium on
  Theory of computing}, pages 206--212, 2003.

\bibitem[MP69]{minsky1969perceptrons}
M.L. Minsky and S.~Papert.
\newblock {\em Perceptrons; an Introduction to Computational Geometry}.
\newblock MIT Press, 1969.

\bibitem[MS22]{malach2022hardness}
Eran Malach and Shai {Shalev-Shwartz}.
\newblock When hardness of approximation meets hardness of learning.
\newblock {\em The Journal of Machine Learning Research}, 23(1):3942--3965,
  2022.

\bibitem[NS94]{nisan1994degree}
Noam Nisan and Mario Szegedy.
\newblock On the degree of boolean functions as real polynomials.
\newblock {\em Computational complexity}, 4(4):301--313, 1994.

\bibitem[Pat92]{paturi1992degree}
Ramamohan Paturi.
\newblock On the degree of polynomials that approximate symmetric boolean
  functions.
\newblock In {\em Proceedings of the twenty-fourth annual ACM symposium on
  Theory of computing}, pages 468--474, 1992.

\bibitem[Rey20]{reyzin2020statistical}
Lev Reyzin.
\newblock Statistical queries and statistical algorithms: Foundations and
  applications.
\newblock {\em arXiv preprint arXiv:2004.00557}, 2020.

\bibitem[RS10]{razborov2010sign}
Alexander~A Razborov and Alexander~A Sherstov.
\newblock The sign-rank of ac$^0$.
\newblock {\em SIAM Journal on Computing}, 39(5):1833--1855, 2010.

\bibitem[Sha18]{shamir2018distribution}
Ohad Shamir.
\newblock Distribution-specific hardness of learning neural networks.
\newblock {\em The Journal of Machine Learning Research}, 19(1):1135--1163,
  2018.

\bibitem[She08a]{sherstov08communicationlower}
Alexander~A. Sherstov.
\newblock Communication lower bounds using dual polynomials.
\newblock {\em Bulletin of the EATCS}, 2008.

\bibitem[She08b]{sherstov2008halfspace}
Alexander~A Sherstov.
\newblock Halfspace matrices.
\newblock {\em Computational Complexity}, 17:149--178, 2008.

\bibitem[She11]{sherstov2011pattern}
Alexander~A Sherstov.
\newblock The pattern matrix method.
\newblock {\em SIAM Journal on Computing}, 40(6):1969--2000, 2011.

\bibitem[She13]{sherstov2013optimal}
Alexander~A Sherstov.
\newblock Optimal bounds for sign-representing the intersection of two
  halfspaces by polynomials.
\newblock {\em Combinatorica}, 33(1):73--96, 2013.

\bibitem[SSS17]{shalev2017failures}
Shai {Shalev-Shwartz}, Ohad Shamir, and Shaked Shammah.
\newblock Failures of gradient-based deep learning.
\newblock In {\em International Conference on Machine Learning}, pages
  3067--3075, 2017.

\bibitem[ST17]{servedio2017circuit}
Rocco~A Servedio and Li-Yang Tan.
\newblock What circuit classes can be learned with non-trivial savings?
\newblock In {\em 8th Innovations in Theoretical Computer Science Conference
  (ITCS 2017)}. Schloss Dagstuhl-Leibniz-Zentrum fuer Informatik, 2017.

\bibitem[STT12]{servedio2012attribute}
Rocco Servedio, Li-Yang Tan, and Justin Thaler.
\newblock Attribute-efficient learning andweight-degree tradeoffs for
  polynomial threshold functions.
\newblock In {\em Conference on Learning Theory}, pages 14--1, 2012.

\bibitem[SW19]{sherstov2019near}
Alexander~A Sherstov and Pei Wu.
\newblock Near-optimal lower bounds on the threshold degree and sign-rank of
  ac0.
\newblock In {\em Proceedings of the 51st Annual ACM SIGACT Symposium on Theory
  of Computing}, pages 401--412, 2019.

\bibitem[Val84]{valiant1984theory}
Leslie~G Valiant.
\newblock A theory of the learnable.
\newblock {\em Communications of the ACM}, 27(11):1134--1142, 1984.

\bibitem[Vap00]{vapnik2000nature}
Vladimir~N. Vapnik.
\newblock {\em The nature of statistical learning theory}.
\newblock Statistics for Engineering and Information Science. Springer-Verlag,
  New York, second edition, 2000.

\end{thebibliography}
\bibliographystyle{alpha}

\end{document}